\documentclass[10pt,a4paper]{article}

\title{Covariance and meaning}

\author{Llu\'{\i}s\ Bel\\
\emph{wtpbedil@lg.ehu.es}
}
\begin{document}

\maketitle

\begin{abstract} 

Several other factors, besides the intrinsic local geometry, contribute to give a meaning to a space-time model. The simplest example comes from comparing Minkowski's and Milne's model, that both have a null Riemann tensor.  We add to these two models a third one which describes a time-dependent locally-Minkowskian spherically symmetric space-time on which every test-particle at rest with respect to the center of symmetry sustains a constant force. Although the model is globally grossly un-realistic we think that it can be helpful to describe a local perturbation of an homogeneous cosmological model. Or as a substitute to the very far away asymptotic Minkowskian behavior usually assumed to describe the gravitational field of compact spherical bodies.

\end{abstract}

\section*{Introduction}

The main purpose of this paper is to get some insight about the meaning of one of the main conceptual ingredients of general relativity: the principle of general covariance. For a majority of the relativity community the principle of general covariance means now that if the Riemann tensor is zero then there is no gravitational field and if it is not zero then there is one. We believe that this over-simplification is a mistake.

	The simplest example that proves that this point of view is wrong is provided by Milne's model of the Universe. This model is too simple to fit what we know today about our Universe, but the fascinating thing is that in the past it was an acceptable model and this suffices to say that it makes sense, and yet the Riemann tensor is zero. This proves that besides the Riemann tensor other factors contribute to give a meaning to a space-time model; and among these factors we include the specification of the differentiable manifold where the metric is regular, the specification of a global frame of reference, and the number and meaning of the free parameters, both essential and non essential ones.
	
The first section is a short review of how one may derive Milne's model emphasizing the role of a distinguished global frame of reference whose time-like congruence is a family of geodesics emanating from a single event of Minkowski's space-time.

The next section deals with a generalization of Milne's model where the distinguished global frame of reference of Milne's model is replaced by one whose time-like world-lines have constant intrinsic curvature. This construction, which takes into account the three dimensions of space, follows similar lines to those followed by M{\o}ller \cite{Moller}, Rindler \cite{Rindler0}, Bel \cite{Bel} and Huang and Sun \cite{Huang}. Our model, whose Riemann tensor is again zero, has two un-essential parameters. It is neither isotropic nor homogeneous and the single symmetry that is easily recognized is the spherical symmetry about a center. Again the point is not whether this model may be useful or not to describe a piece of real physics. What is important to us is that it could be because it has a meaning and this meaning is different from Minkowski's and Milne's models.

The third section deals with light propagation along radial directions toward and away from the center of symmetry
and the derivation of the red-shift formulas in a variety of cases.

In the fourth section we linearize the preceding results by considering an approximation around the present epoch and taking into account an appropriate span of distances. And in the fifth section we describe a very simple illustrative example that shows how our model could be falsified.
 
Our concluding remarks review some of the main points mentioned in this introduction, that the main body of the paper clarifies. We hope that our model will be useful. If not as it stands at least as a motivation for further work along a similar direction.

%%%%%%%%%%%%%%%%%%%%%%%%%%%%%%%%%%%%Milne's model of the Universe

\section{Milne's model of the Universe}

Let us consider an event $x_0$ of Minkowski's space-time model: 

\begin{equation}
\label{0.1}
ds^2=-dt^2+\delta_{ij} dx^idx^j,\ \hbox{or}\ ds^2=-dt^2+dr^2+r^2d\Omega^2                       
\end{equation}
in Cartesian $(x^i)$ or spherical coordinates $(r,\ \theta,\ \varphi)$. 

Let:

\begin{equation}
\label{0.2}
t=\cosh(\alpha)\tau, \ \ x^i=\lambda^i\sinh(\alpha)\tau                      
\end{equation}
be the congruence of future pointing time-like geodesics originated at $x_0$, with coordinates $t_0=0,\ x_0^i=0$, where  $\tau$ is the evolution proper-time along each of these  geodesics, $\lambda^i$ are the components of a unit space vector, and $\alpha\in [0,\infty)$ is the canonical parameter of a special Lorentz transformation. $\lambda^i\alpha$ are dimensionless constants and therefore they can not be considered directly as coordinates of a physical space. This can be cured\,\footnote{This is not usually done. See Refs. \cite{McVittie}, \cite{Rindler}} if we introduce a free parameter, say $p>0$, with dimension [1/length] and define:

\begin{equation}
\label{0.3}
R=\frac{\alpha}{p}, \quad z^i=R\lambda^i                 
\end{equation} 
so that in spherical coordinates the congruence defined in (\ref{0.2}) is:

\begin{equation}
\label{0.4}
t=\cosh(pR)\tau,\ r=\sinh(pR)\tau , \ \theta=\theta, \ \varphi=\varphi                       
\end{equation} 
Using the adapted coordinates $(\tau,R,\theta,\varphi)$ the line element (\ref{0.1}) becomes:

\begin{equation}
\label{0.9}
ds^2=-d\tau^2+p^2\tau^2(dR^2+p^{-2}\sinh(pR)^2d\Omega^2)                     
\end{equation}
which is the line-element of Milne's model of space-time using our preferred radial coordinate\,\footnote{In \cite{McVittie} and \cite{Rindler} a different radial coordinate is used} $R$.

Minkowski's and Milne's models are both locally flat. And both have privileged frames of reference with time-like world-lines that are geodesics. Why are then different models of universes? Because the concept of relative rest has changed. Let us consider two free test particles. In Minkowski's universe the two particles are at relative rest if the two geodesic world-lines are parallel. In Milne's model the two particles are at relative rest if the two geodesics intersect at one prescribed event.

In the model described in the next section the concept of relative rest is again different.
 
%%%%%%%%%%%%%%%%%%%%%%%%%%%%%%%%%%% Generalization of Milne's model
 
\section{Generalization of Milne's model}

Let us consider an event $x_0$ of Minkowski's space-time model as in the preceding section  and let $\vec e_\alpha$ be an orthonormal frame of reference at $x_0$.

Let us consider a time-like world-line of constant curvature outgoing from $x_0$ and being tangent to $\vec e_0$. Its parametric equations can be written:

\begin{equation}
\label{1.2}
x^i=\frac{1}{a}\lambda^i(\cosh(a\tau)-1), \quad  t=\frac{1}{a}\sinh(a\tau)
\end{equation}
where $a$ is the constant acceleration:

\begin{equation}
\label{1.3}
a^2=\delta_{ij}\ddot x^i \ddot x^j-\ddot t^2
\end{equation}
a dot meaning a derivative with respect to the parameter $\tau$, and $\lambda^i$ being the unit vector in the  direction of the space velocity: 

\begin{equation}
\label{1.4.1}
\dot x^i=\lambda^i\sinh(a\tau), \quad \dot t=\cosh(a\tau), \quad \delta_{ij}\dot x^i \dot x^j-\dot t^2=-1
\end{equation}
as well as the direction of the space acceleration:

\begin{equation}
\label{1.4}
\ddot x^i=a\lambda^i\cosh(a\tau), \quad \ddot t=a\sinh(a\tau)
\end{equation}
 
With varying values of $\lambda^i$ we obtain thus a 2-parameter congruence that can be extended to a 3-parameter one considering all the world lines that can be derived from any of the preceding ones by a special Lorentz transformation with canonical parameter $\alpha$. We obtain then a congruence $\cal C$ defined by the parametric equations:

\begin{eqnarray}
\label{1.5}
x^i&=&\frac{\lambda^i}{a}\left(\cosh(\alpha)(\cosh(a\tau)-1)+\sinh(\alpha)\sinh(a\tau)\right) \\
\label{1.5.1}
t&=&\frac{1}{a}\left(\cosh(\alpha)\sinh(a\tau)+\sinh(\alpha)(\cosh(a\tau)-1)\right)
\end{eqnarray}
or equivalently:

\begin{eqnarray}
\label{1.5.2}
x^i&=&\frac{\lambda^i}{a}(\cosh(\alpha+a\tau)-\cosh(\alpha)) \\
\label{1.5.3}
t&=&\frac{1}{a}(\sinh(\alpha+a\tau)-\sinh(\alpha))
\end{eqnarray}
Defining as in (\ref{0.3}) the adapted coordinates $z^i$ and $R$,
Eqs. (\ref{1.5}) and (\ref{1.5.1}) become:

\begin{eqnarray}
\label{1.8}
x^i&=&\frac{z^i}{aR}(\cosh(pR+a\tau)-\cosh(pR)) \\
\label{1.8.1}
t&=&\frac{1}{a}(\sinh(pR+a\tau)-\sinh(pR))
\end{eqnarray}
and from the first of these we get:

\begin{equation}
\label{1.9}
r=\frac{1}{a}(\cosh(pR+a\tau)-\cosh(pR))
\end{equation}
so that using spherical coordinates $(R,\ \theta,\ \varphi)$ we obtain the following model of space-time locally homeomorphic with Minkowski's one:

\begin{equation}
\label{1.13}
dS^2=-d\tau^2+\frac{2p^2}{a^2}(\cosh(a\tau)-1)dR^2+\frac{2p}{a}(\cosh(a\tau)-1)dRd\tau+
r^2d\Omega^2
\end{equation}

This line-element is neither isotropic nor homogeneous. It is time dependent and spherically symmetric, the center of symmetry being the world-line $R=0$, and its limit when the parameter $a$ tends to zero is Milne's line-element (\ref{0.9})

Let $u$ be the unit time-like vector field tangent to the world-lines with constant $R,\ \theta,\ \varphi$ coordinates and variable $\tau$. Its dual 1-form is:

\begin{equation}
\label{1.14}
\psi^0=-d\tau+\frac{p}{a}(\cosh{a\tau}-1)dR                   
\end{equation}
that together with:

\begin{equation}
\label{1.15}
\hat\psi^1=\frac{p}{a}\sinh(a\tau) dR,\ \hat\psi^2=rd\theta,\ \hat\psi^3=r\sin\theta d\varphi                   
\end{equation}
completes an orthonormal decomposition of (\ref{1.13}):

\begin{equation}
\label{1.16}
dS^2=-(\psi^0)^2+(\hat\psi^1)^2+(\hat\psi^2)^2+(\hat\psi^3)^2                  
\end{equation}
We have:

\begin{equation}
\label{1.17}
d\psi^0=p\sinh(a\tau)d\tau\,\wedge\, dR=-a\psi^0\,\wedge\, \hat\psi^1               
\end{equation}
from where it follows that:

\begin{equation}
\label{1.18}
d\psi^0\,\wedge\, \psi^0=0, \ i(u)\,d\psi^0=-a\hat\psi^1             
\end{equation}
where $\wedge$ is the exterior product symbol, $i(u)$ the interior product one and $d$ the exterior differential operator.

The first result means that the congruence defined by $u$ is orthogonal to a family of hypersurfaces, that can be proved to be:

\begin{equation}
\label{1.19}
U=\exp(pR)\exp\left(\tanh\left(\frac{a\tau}{2}\right)^{-1}\right)=const.            
\end{equation}
The use of a time coordinate proportional to $U$ would diagonalize the line-element (\ref{1.13}) but we shall not use it.

The second result (\ref{1.18}) means that the intrinsic curvature for every world-line of the congruence is $a$ in the radial direction, a property that was built in the model of space-time from the very beginning. Equivalently, we can see using the geodesic equations that when a free falling has zero velocity at some $R=R_1$, then:

\begin{equation}
\label{1.19.1}
\left(\frac{d^2R}{d\tau^2}\right)_1=-\frac{a^2}{p}\sinh(a\tau)^{-1}           
\end{equation}

The line-element (\ref{1.13}) is regular for any value of $\tau\ne 0$ and any value of $R$. But it has a defaults at $R=0$: the coefficient of $d\Omega^2$ is:

\begin{equation}
\label{1.20.1}
r^2=\frac{1}{a^2}(\cosh(a\tau)-1)^2            
\end{equation}
and therefore $R=0$ can not be identified with a point.

To solve this difficulty we could generalize this model using instead of a constant $a$ an appropriate function $a(R)$. One of the simplest would be:

\begin{equation}
\label{1.21.2}
a(R)=\frac{aR^2}{b^2+R^2}           
\end{equation}
where $b$ is a new parameter with dimensions [length]. With this function we should have for small enough values of $R$:

\begin{equation}
\label{1.22.2}
a(R)=\frac{a}{b^2}R^2         
\end{equation}
and for large values of $R$ we should recover the line-element(\ref{1.13}).

More precisely the line-element that we should obtain for small values of $R$ is:

\begin{equation}
\label{1.23.2}
dS^2=-d\tau^2+p\tau^2\left(p+2\frac{a}{b^2}\tau R\right)dR^2
+\frac{pa}{b^2}\tau^2R^2dRd\tau+p\tau^2\left(p+\frac{a}{b^2}\tau R\right)R^2d\Omega^2  
\end{equation}

Our initial model therefore may be considered as being asymptotic to a more elaborate one that in fact we shall not need in the sequel of this paper.

%%%%%%%%%%%%%%%%%%%%%%%%%%%%%%%%%%%%% Radial light propagation

\section{Radial light propagation}

Let us consider two points with coordinates ($R_1,\ \theta_1,\ \varphi_1$) and ($R_2,\ \theta_2,\ \varphi_2$). We consider three cases: in the first case (i) we assume that $R_1>R_2,\ \theta_1=\theta_2,\ \varphi_1=\varphi_2$;  in the second case (ii) we assume that $R_1<R_2,\ \theta_1=\theta_2,\ \varphi_1=\varphi_2$; and in the third case (iii) we assume that $\theta_1=\pi-\theta_2\ \varphi_1=\varphi_2+\pi$.

 When light travels from $R_1$ to $R_2$ its coordinate speed:

\begin{equation}
\label{1.20}
V=\frac{dR}{d\tau} 
\end{equation}
must be a solution of the quadratic equation:

\begin{equation}
\label{1.21}
-1+\frac{2p^2}{a^2}(\cosh(a\tau)-1)V^2+\frac{2p}{a}(\cosh(a\tau)-1)V=0
\end{equation}
The two solutions of this equation are:

\begin{equation}
\label{1.22}
V_1=-\frac{a\exp(a\tau)}{p(\exp(a\tau)-1)},\ \ V_2=\frac{a}{p(\exp(a\tau)-1)},\ 
\end{equation}
with a mean value of the speed:

\begin{equation}
\label{1.22.1}
V_m=\frac12(V_2-V_1)=\frac{a}{2p}\frac{\exp(a\tau)+1}{\exp(a\tau)-1}
\end{equation}

In case (i) above $R$ decreases when $\tau$ increases and therefore, assuming that $a$ and $p$ are both positives, we have $V=V_1$, while in case (ii) we have $V=V_2$. In case (iii) we have $V=V_1$ when light goes from $R_1$ to the center of symmetry $R=0$ and $V=V_2$ when light travels from the center of symmetry to $R_2$

Integrating the Eq. (\ref{1.20}) from $\tau_1$ to $\tau_2$ in the case (i) we have:

\begin{equation}
\label{1.23}
R_1-R_2=-\int_{\tau_1}^{\tau_2}\frac{a\exp(a\tau)\,d\tau}{p(\exp(a\tau)-1)}=\frac{1}{p}\ln\frac{\exp(a\tau_2)-1}{\exp(a\tau_1)-1}
\end{equation}
or:

\begin{equation}
\label{1.23.1}
\exp(p(R_1-R_2))=\frac{\exp(a\tau_2)-1}{\exp(a\tau_1)-1}.
\end{equation}
Differentiating this equation with respect to $\tau_1$, keeping fixed $R_1-R_2$, we obtain:

\begin{equation}
\label{1.24.1}
\frac{d\tau_2}{d\tau_1}=\frac{(\exp(a\tau_2)-1)(\exp(a\tau_1)}{(\exp(a\tau_1)-1)(\exp(a\tau_2)};
\end{equation}
and solving this same equation (\ref{1.23.1}) for $\tau_1$ we get:

\begin{equation}
\label{1.24}
\exp(a\tau_1)=\frac{\exp(p(R_1-R_2))-1+\exp(a\tau_2)}{\exp(p(R_1-R_2))}
\end{equation}
We obtain thus substituting this expression in the above equation (\ref{1.24.1}):

\begin{equation}
\label{1.24.2}
z=\exp(-a\tau_2)\left(\exp(p(R_1-R_2))-1\right)
\end{equation}
where:

\begin{equation}
\label{1.24.3}
z=\frac{d\tau_2}{d\tau_1}-1
\end{equation}
is the red-shift of the light emitted from $R_1$ at time $\tau_1$ when received at $R_2$ at time $\tau_2$.

A similar calculation in the case (ii) leads to the following results:

\begin{equation}
\label{1.26}
z=\exp(a\tau_2)(\exp(p(R_2-R_1))-1)
\end{equation}

Finally, to calculate the red-shift in the case (iii) we proceed as follows: we use (\ref{1.24.1}) with $\tau_2=\tau^*$ where $\tau^*$ is the arrival time at the center of symmetry $R=0$. Then we use the corresponding formula in the case (ii) with $\tau_1=\tau^*$; we multiply both expressions and write the result as a function of the arrival time only, getting:

\begin{eqnarray}
&& z=\frac{\exp(pR_1)(\exp(pR_2+a\tau_2)+1-\exp(a\tau_2))^2}{\exp(pR_2+a\tau_2)} \nonumber\\
\label{1.27}
&& \hspace{1cm}-\frac{(1-\exp(a\tau_2))(\exp(pR_2+a\tau_2)+1-\exp(a\tau_2))}{\exp(pR_2+a\tau_2)}-1
\end{eqnarray}

%%%%%%%%%%%%%%%%%%%%%%%%%%%%%%%%%% The linear approximation}

\section{The linear approximation}

We proceed in this section to linearize the line-element (\ref{1.13}) around a prescribed epoch $\tau_0$. Let us change the origin of the time coordinate so that:

\begin{equation}
\label{3.1}
\tau=T+\tau_0
\end{equation}
Keeping only terms linear in $aT$ we get the following approximation:

\begin{equation}
\label{3.2}
dS^2=dS^2_0+dS^2_1
\end{equation}
where:

\begin{equation}
\label{3.3}
dS^2_0=-\left(dT+\frac{p}{a}(\cosh(a\tau_0)-1)dR\right)^2+\frac{p^2}{a^2}\sinh(a\tau_0)^2dR^2+r_0^2d\Omega^2,
\end{equation}
with:

\begin{equation}
\label{3.3.1}
r_0=\frac{1}{a}(\cosh(pR+a\tau_0)-\cosh(pR))
\end{equation}
and:
\begin{equation}
\label{3.4}
dS^2_1=2\sinh(a\tau_0)pTdRdT+2\frac{p}{a}\sinh(a\tau_0)pTdR^2+2r_0r_1d\Omega^2
\end{equation}
with:

\begin{equation}
\label{3.5}
r_1=\sinh(pR+a\tau_0)T
\end{equation}

With the linear coordinate transformation:

\begin{equation}
\label{3.10}
T\leftarrow T+\frac{p}{a}(\cosh(a\tau_0)-1)R
\end{equation} 
we obtain:

\begin{equation}
\label{3.11}
dS^2_0=-dT^2+\frac{p^2}{a^2}\sinh(a\tau_0)^2dR^2+r_0^2d\Omega^2,
\end{equation}
and :

\begin{eqnarray}
&& 
\hspace{-1cm} dS^2_1=\frac{2p}{a}\sinh(a\tau_0)\left(aT+pR(\cosh(a\tau_0)-1)\right)dRdT \nonumber \\
\label{3.11.0}
&& 
\hspace{-.5cm} +\frac{2p^2}{a^2}\sinh(a\tau_0)\cosh(a\tau_0)\left(aT+pR(\cosh(a\tau_0)-1)\right)dR^2+2r_0r_1d\Omega^2
\end{eqnarray}

The next approximation that can be useful to consider will be to keep only linear terms of $pR$. In which case Eqs.  and (\ref{3.3.1})and (\ref{3.5})become: 

\begin{eqnarray}
\label{3.11.1}
r_0&=&\frac1a(\cosh(a\tau_0)-1+\sinh(a\tau_0)pR)\\
r_1&=&(\sinh(a\tau_0)+\cosh(a\tau_0)pR)T
\end{eqnarray}
and the red-shift formulas of the preceding section:

\begin{equation}
\label{3.12.1}
z=\exp(-a\tau_0)p(R_1-R_2)
\end{equation}
in the case (i);

\begin{equation}
\label{3.12.2}
z=\exp(a\tau_0)p(R_2-R_1)
\end{equation}
in the case (ii); and:

\begin{equation}
\label{3.12.3}
z=\exp(a\tau_0)pR_2+\exp(-a\tau_0)pR_1
\end{equation}
in the case (iii), where $\tau_0$ is the epoch at the reception time of the light.

Notice that at this approximation $p$ plays the role of the Hubble constant as in Milne's model but with this generalized model this constant is modulated by factors that depend on the epoch and the case which is considered. 

%%%%%%%%%%%%%%%%%%%%%%%%%%%%%%%%%%%%%%%%%%%%%Examle

\section{Example}

To figure out what sort of information our model could provide to falsify it we consider below a very speculative example. 

We imagine that the Solar system, as any other concentration of matter, can locally perturb the background cosmological Universe, or modify our description of it, to a degree that depends of its mass. More precisely we assume below that our model is adequate to deal with either case assuming that the Sun is at the center of symmetry. We ignore for simplicity the gravitational field of the Sun, but otherwise we use local physics as we know at the present epoch.

In this case Eqs. (\ref{3.11}) and  (\ref{1.22.1}) suggest us to define the present epoch by the value of $a\tau_0$ such that:

\begin{equation}
\label{3.11.2}
\frac{p}{a}=1,\quad \frac12(V_2-V_1)=\frac{a}{2p}\frac{\exp(a\tau_0)+1}{\exp(a\tau_0)-1}=1 
\end{equation}
These conditions yield the following results:

\begin{equation}
\label{3.14}
a\tau_0=1.1
\end{equation}

From (\ref{1.24.2}) we see that for a solar system observer the Hubble law will be:

\begin{equation}
\label{3.13.1}
z=0.33\,pR
\end{equation}
and from (\ref{1.19.1}) it follows that he will observe an ``anomalous'' acceleration of free falling particles toward the center given by:

\begin{equation}
\label{3.14.1}
\frac{d^2R}{dT^2}=-0.75\,a
\end{equation}  

\section*{Concluding remarks}

Minkowski's, Milne's and the model that we have described in this paper have a common property: the Riemann tensor is zero, and yet each has a different meaning. Minkowski's model is totally unacceptable as a description of an evolving universe while Milne's model is adequate to describe the basic fact of Cosmology, i.e the feature which is usually called the expansion of the Universe; and our model will or will not have a future in Cosmology but what is important is that it could have one, introducing new features that had not been anticipated until recently: as the acceleration of the expansion of the Universe or the manifestation of anomalous forces, as suggested by the so-called Pioneer's anomaly.

The variety of meanings depends on three main factors. The first one is the manifold that supports the Riemannian metrics. The second is the number and the role of distinguished frames of reference, from which depend the concept of relative rest. And the third is the number, the role and of course the value of the parameters on which depend the corresponding line-elements. 

The differential manifolds $V_4$ of the three models are different. In Minkowski's model $V_4=R^4$; in both Milne's and our model $V_4$ is homeomorphic to the interior of any future-pointing null cone of Minkowski's space-time.

Minkowski's model has a 3-parameter family of distinguished global frames of reference: the Galilean ones. On the contrary, Milne's and our model have each a single distinguished global frame of reference. The corresponding time-like congruences are a family of geodesics for Milne's model and a family of constant acceleration world-lines for our model.  
  
Minkowski's model does not contain any parameter. Milne's model contains one that we have called p and our model contains two: p and a. Or three if the the neighborhood of $R=0$ is regularized as indicated at the end the Sect. 2. They are ``un-essential´´ locally because they can be eliminated with the coordinate transformation inverse to that that we used to derive them.  But they are essential in the sense that the meaning of each model depends on them. If $p=0$ then Milne's model is Minkowski's model. If $a=0$ then our model is Milne's model.

The role of $p$ in deriving Milne's and our model is crucial. In both cases the quantity $\alpha$ is an adapted coordinate of the corresponding distinguished frame of reference but it is a dimensionless quantity, when a space coordinate should have dimension [length]. This is the reason why the formula (\ref{0.3}), which introduces the parameter $p$ with dimension [1/length], and defines the variable $R$ is necessary. The role of $a$ is also  crucial because the new features of our model depend on it.

Our model is not isotropic nor homogeneous and has a center. Therefore it can not describe a real Universe which is isotropic and does not have anything that we could call its center. But our real Universe certainly does not have these properties at any scale, even if we still can assume that there is a scale at which it does have these properties. Therefore our model could still be useful to describe large perturbations of isotropy or homogeneity as they could be caused by the Great attractor or the Shapley concentration. Or it could maybe be useful also as a substitute to the Minkowskian behavior at large distances of those gravitational fields with central sources. 

\section*{Acknowledgements}

A. Molina has friendly contributed to improve a first version of this manuscript.

\end{document}